\documentclass[preprint2,numberedappendix]{emulateapj-rtx4}

\usepackage{amssymb,bm}
\usepackage{epsfig}
\usepackage{natbib}

\newcommand{\beq}{\begin{eqnarray}}
\newcommand{\eeq}{\end{eqnarray}}

\newcommand{\ff}{\bm{f}}
\newcommand{\DD}{{\rm{D}}}
\newcommand{\bra}[1]{\langle #1\rangle}
\newcommand{\nab}{\mbox{\boldmath $\nabla$} {}}
\newcommand{\Sec}[1]{Section~\ref{#1}}
\newcommand{\Fig}[1]{Figure~\ref{#1}}

\shorttitle{Magnetic Fields from QCD Phase Transitions}
\shortauthors{Tevzadze et al.}

\begin{document}

\title{Magnetic Fields from QCD Phase Transitions}

\date{}


\author{
Alexander G. Tevzadze$^{1}$,
Leonard Kisslinger$^2$,
Axel Brandenburg$^{3,4}$,
and
Tina Kahniashvili$^{2,5,6}$
\email{aleko@tevza.org
($ $Revision: 1.62 $ $)}
}

\affil{ $^1$Faculty of Exact and Natural Sciences, Javakhishvili
    Tbilisi State University, 1 Chavchavadze Ave., Tbilisi, 0128, Georgia\\
$^2$McWilliams Center for Cosmology and Department of Physics, Carnegie Mellon University,
    5000 Forbes Ave, Pittsburgh, PA 15213\\
$^3$Nordita, KTH Royal Institute of Technology and Stockholm University,
    Roslagstullsbacken 23, 10691 Stockholm, Sweden\\
$^4$Department of Astronomy, AlbaNova University Center, Stockholm University,
    10691 Stockholm, Sweden\\
$^5$Department of Physics, Laurentian, University, Ramsey Lake Road, Sudbury, ON P3E 2C, Canada\\
$^6$Abastumani Astrophysical Observatory, Ilia State University,
    3-5 Cholokashvili Str., Tbilisi, 0194, Georgia
}

\begin{abstract}

We study the evolution of QCD phase transition-generated magnetic
fields in freely decaying MHD turbulence of the expanding Universe.
We consider a magnetic field generation model that starts from basic
non-perturbative QCD theory and predicts stochastic magnetic fields
with an amplitude of the order of 0.02\,$\mu$G and small magnetic
helicity. We employ direct numerical simulations to model the MHD
turbulence decay and identify two different regimes: ``weakly
helical'' turbulence regime, when magnetic helicity increases during
decay, and ``fully helical'' turbulence, when maximal magnetic
helicity is reached and an inverse cascade develops. The results of our
analysis show that in the most optimistic scenario the magnetic
correlation length in the comoving frame can reach 10\,kpc with the
amplitude of the effective magnetic field being 0.007\,nG. We
demonstrate that the considered model of magneto-genesis can provide
the seed magnetic field for galaxies and clusters.

\end{abstract}


\maketitle

\section{Introduction}

The origin of the observed magnetic fields (MFs) in galaxies and
clusters of $\sim 10^{-6}$ -- $10^{-5}$ Gauss (G) remains a matter
of debate \citep{Beck,Widrow,Vallee}. Recently several different
groups \citep{neronov,limit2,dolag2,tay,huan} reported the detection
of a lower bound on a large-scale correlated MF amplitude of the
order of $10^{-16}-10^{-15}$ G, or possibly two orders of magnitude
smaller \citep{dermer,tak} at Mpc scales through blazar
observations. One of the possible explanations of the large-scale
correlated MF assumes the presence of a seed primordial magnetic
field (PMF) which was generated during or prior to the radiation
dominated epoch. This MF should satisfy several conditions: (i) The
PMF should preserve approximate spatial isotropy, it has to be weak
enough when its energy density can be treated as a first order of
perturbation; (ii) The PMF should be smaller than the MF in galaxies
by a few orders of magnitude at least, since during structure
formation PMFs get amplified; (iii) Since the PMF energy density
$\rho_B$ contributes to the radiation field, the big bang
nucleosynthesis (BBN) bound implies $\Omega_B h_0^2
=\rho_B/\rho_{\rm cr} \leq 2.4\times 10^{-6}$ \citep{grasso}, where
$\rho_{\rm cr}$ is the critical density,\footnote{The ratio of
$\rho_B$ to the energy density of the radiation $\rho_{\rm rad}$ is
constant during cosmological evolution if the PMF is not damped by a
MHD (or other) process and therefore stays frozen into the plasma.}
and $h_0$ is the Hubble constant in units of 100 km s$^{-1}$
Mpc$^{-1}$.

The possible origin of the PMF from the two major cosmological phase
transitions, the electroweak phase transition (EWPT) and the QCD
phase transition (QCDPT) \citep[see][for
reviews]{grasso,Widrow,review2012} is of particular importance for
cosmology. Because of the larger scale of the resulting seed
magnetic field and the nature of the QCD bubble walls during a first
order QCDPT, it is more likely that the QCDPT rather than the EWPT
produces a PMF that accounts for the observed galactic and cluster
MFs.

In this paper we consider one of several possible mechanisms of PMF
generation. In particular we re-address the model proposed by
\cite{kisslinger}, in which the PMF is generated via QCD bubble
collisions. We consider the coupling of this initial PMF with the
QCD plasma, and study the dynamics during the expansion of the
Universe. The main parameters of the described model are given by
the QCDPT temperature $T_\star=0.15$ GeV and the number of
relativistic degrees of freedom $g_\star =15$. The interactions
between the PMF and the QCD plasma is studied through numerical MHD
simulations using the {\sc Pencil Code} (see
http://pencil-code.googlecode.com/). We discuss observational
signatures of such a QCDPT PMF, including observed MFs in galaxies
and clusters. We employ natural units with $\hbar = 1 = c$ and
gaussian units for the MHD formulation.

The outline of the paper is as follows: In Section 2 we describe the
PMF generation model. In Section 3 we determine the spatial and
temporal characteristics of the generated PMF. The results of our
analysis, including the dynamics of the PMF, are presented in
Section 4, where we discuss the resulting MF in galaxies and
clusters. Conclusions are presented in Section 5.

\section{Magnetic Field Generation Model}

In contrast to the EWPT, the QCDPT involves the treatment of QCD,
which, unlike the electroweak theory, is non-perturbative. Therefore a
valid theory starting from basic QCD theory, rather than a model,
must be able to treat non-perturbative QCD. In \cite{kisslinger},
instantons form gluonic bubble walls and it is the interior
gluonic wall that leads to the magnetic seed described below
satisfy that criterion. In this early work the main interest was
the prediction of polarization correlations in cosmic microwave
background radiation (CMBR). As we can see below the magnitude of
the resulting MF is too small for current CMBR observations, but it
might be measured in the future. Because this scenario starts from
basic non-perturbative QCD theory and successfully predicts a primary
magnetic field which has the overall properties that are promising
for the PMF, we use it in our present work.

In this section we briefly describe the PMF scenario proposed by
\cite{kisslinger}. In \Sec{During} we review the magnetic
field and helicity density created during the QCDPT, and in \Sec{Comoving}
we give values of these quantities at the present time.
Recent lattice QCD studies have shown that  the QCDPT is first
order, so bubbles form and collide \citep[see][and references
therein]{qcd1,qcd2,qcd3,qcd4,qcd,Boeckel:2011y}; additional
references are given by \cite{kks10}. The first order QCDPT can
result in the generation of a MF through two (or more) bubble
collisions.

\subsection{Magnetic Field and Helicity During the QCDPT}
\label{During}

The QCD phase transition critical temperature
is defined as $T_\star \simeq $ 0.15 GeV. A gluonic wall is created
as two bubbles collide, and a magnetic wall is formed by the
interaction of the nucleons with the gluonic wall.
The electromagnetic interaction Lagrangian is
\begin{equation} \label{Lint}
{\mathcal L}^{\rm int} =  -e \bar{\Psi} \gamma^\mu A^{\rm em}_\mu \Psi ~,
\end{equation}
where $\Psi$ is the nucleon field operator, $A^{\rm em}$ is the
electromagnetic 4-potential, and $\gamma^\mu$ are the Dirac
matrices. In \cite{kisslinger} it was shown that the interaction of
the quarks in the nucleons with the gluonic wall align the nucleon
magnetic dipole moments, producing a $B$-field orthogonal to the
gluonic wall.

Using an instanton model for the gluonic wall oriented in the $x$-$y$
direction (say), one obtains for $B_z({\bm x})$ at the time of the QCDPT,
with $T=T_\star$,
\begin{equation} \label{Bx}
B_z({\bm x}) = B^{(\rm QCD)}_\star e^{-b^2(x^2 + y^2)} e^{-M_n^2 z^2} ~,
\end{equation}
where $b^{-1}=d_H \simeq$ a few km =horizon size at the end of the QCDPT
($t\simeq 10^{-4}$s) and $M_n^{-1}$ = 0.2 fm.
$B^{\rm QCD}_\star$, the magnitude of the MF within the wall of
thickness $\zeta$, is \citep[see][]{kisslinger}
\begin{equation} \label{bz}
B^{(\rm QCD)}_\star  \simeq \frac{1}{\zeta\Lambda_{\rm QCD}}
\frac{e}{2 M_n} \times \langle \bar{\Psi} \sigma_{21}\gamma_5 \Psi
\rangle ~,
\end{equation}
where $\Lambda_{\rm QCD} \simeq 0.15$ GeV is the QCD momentum scale,
$\gamma_5 =i\gamma^0 \gamma^1 \gamma^2 \gamma^3$, and $\sigma_{21}=
i \gamma_2 \gamma_1 =i \gamma^2 \gamma^1$. A similar form had been
derived earlier using the domain wall model of \cite{fz00}. The value
for $B^{(\rm QCD)}$ was found to be
\begin{eqnarray} \label{bw}
B^{(\rm QCD)}_\star \simeq 0.39 \frac{e}{\pi} \Lambda_{\rm QCD}^2 \simeq
1.5 \times 10^{-3}~ {\rm GeV^2} \nonumber \\
\simeq 2.2 \times 10^{16} ~{\rm G} ~.
\end{eqnarray}
The asterisk indicates that we refer to the {\it initial} value
of the MF at the time of the QCDPT.

We now discuss the magnetic helicity created during the QCDPT using
the scenario proposed by \cite{kisslinger}. Magnetic helicity is an
important characteristic that strongly influences the PMF dynamics.
Magnetic helicity is a conserved quantity during the subsequent evolution
past the QCDPT. This leads to an inverse
cascade producing magnetic fields at progressively larger scales.
For this to work, it is important to know
the magnetic helicity that is produced by the QCDPT.

The magnetic helicity is defined as $\int \! {\rm d}^3 x \, {\textbf A}
\cdot {\bm B}$, with ${\bm B} = \nabla \times {\textbf A}$.
In the domain wall model of \cite{fz00}, the magnetic helicity density
${\mathcal H}_M$ is
\begin{equation} \label{HB}
{\mathcal H}_M = {\bm A} \cdot {\bm B} = A_z B_z ~,
\end{equation}
for a PMF in the $z$ direction, as discussed above.
Because of strong CP violation during the QCDPT, magnetic helicity
is produced through the alignment of magnetic and electric dipole
moments of the nucleons.
Thus, the electric field satisfies $E_z \simeq B_z$ \citep[see][]{fz00}.
From Maxwell's equations in the Weyl gauge we have
\begin{eqnarray} \label{E}
{\bm E} = -\frac{1}{c} \frac{\partial {\bm A}}{\partial t}
\quad\mbox{or}\quad
A_z &\simeq& -E_z \tau ~,
\end{eqnarray}
where $\tau \simeq 1/\Lambda_{\rm QCD}$ is the timescale for the
QCDPT. From Eqs.~(\ref{HB}) and (\ref{E}) one finds
\begin{eqnarray} \label{hz}
{\mathcal H}_{M, \star}^{\rm QCD} &\simeq& B_z^2/\Lambda_{\rm QCD} \nonumber \\
&\simeq& (0.22 \times 10^{17}~{\rm G})^2/(0.15 {\rm GeV}) ~,
\end{eqnarray}
where we have assumed statistical homogeneity, so the result is
gauge-independent.

\subsection{Comoving Values of Magnetic Field and Helicity}
\label{Comoving}

The simple dilation due to the expansion of the Universe
significantly reduces the amplitude of both the MF and the magnetic
helicity created during the QCDPT. Defining $a_\star$ and $a_0$ as
the scale factors at the time of the QCDPT and today,
respectively, we have
\beq
\label{a-ratio} \frac{a_\star}{a_0} &\simeq& 10^{-12}
\left(\frac{0.15\,{\rm GeV}}{T_\star}\right)
\left(\frac{15}{g_\star}\right)^{{1}/{3}} \; ,
\eeq
with $g_*$=15, $T_\star$=0.15 GeV.

The comoving (present) value of the PMF field $B_{\rm in} $ (the subscript
``in'' indicates that the QCD field is an initial PMF for further
developed MHD dynamics) is given by $ B_{\rm in} =
({a_\star}/{a_0})^2 \times B^{(\rm QCD)}_\star$, which results in
\begin{equation}
B_{\rm in} \simeq  2 \times 10^{-8}~{\rm G}.
\label{initial}
\end{equation}

As in the case of the PMF amplitude, magnetic helicity density
experiences dilution due to the expansion of the Universe. The
comoving (initial) value of the magnetic helicity density is given by
\begin{equation}
{\mathcal H}_{M, {\rm in}} = \left(\frac{a_\star}{a_0}\right)^3  \times
{\mathcal H}_{M, \star}^{\rm QCD} \simeq 10^{-39}~({\rm G}^2 \cdot {\rm Mpc}).
\label{weakhel}
\end{equation}
This value is extremely small, and it is almost $\times 10^{19}$
smaller than the maximal allowed magnetic helicity (see below). Such
a small value of magnetic helicity density is due to the thickness
of the magnetic wall; see Eq.~(\ref{Bx}) and \cite{kisslinger}. On
the other hand several studies indicate strong CP violation during
QCDPT \citep{cp2,cp3,cp4}. In this case magnetic helicity can
reach its maximal value, if we assume the MF to be correlated over
the Hubble scale $\lambda_{H_\star}$, as will be explained below.
Being more conservative we assume that the MF correlation should
coincide with the bubble size $\xi_M$, see Sec.~3. The resulting
magnetic helicity will then be smaller than the maximal one by a
factor of the order of $\xi_M/\lambda_{H_\star}$. As we will see in
Sec.~4, the duration of the process is long enough to ensure that
the maximal value of magnetic helicity is reached during the
subsequent evolution.

\section{Magnetic Field Spectrum}

Following our earlier studies \citep{kbtr10} we treat the initial PMF
energy density ${\mathcal E}_M$ as magnetic energy density
injected into the cosmological plasma at the comoving length scale
$\lambda_{0}$ which corresponds to the QCD bubble size. We recall
that the PMF has been generated on the thin surfaces between colliding
bubbles, while the correlation length scale of this PMF might be
associated with the bubble length scale. In the following, we assume
that the PMF spectrum in Fourier space is sharply peaked at
$k_0=2\pi/\lambda_0$. After generation, the PMF evolution (during
the PT) depends sensitively on the length scale under consideration
and on the presence of magnetic helicity \citep[see][for magnetic helicity
generation mechanisms]{pmf1,turner,hel1,Jackiw,Garretson,Field,
Giovannini,hel4,campanelli,hel7,Durrer3}. The expansion of the
Universe leads to additional effects, in particular to a faster
growth of the PMF correlation length. A distinctive effect is the
different time behavior of the PMF decay.

In the cosmological context most important is the difference between
the growth of the comoving length scale ($L \propto a$) and the
Hubble radius ($H^{-1} \propto t$, where $t$ is physical time). This
leads to additional effects in the PMF evolution (and damping)
\citep[see][]{BN992,jedamzik1,Caprini2010}. Note that, to describe
properly the dynamics of the perturbations in the expanding
Universe, it is appropriate to switch to comoving quantities and to
describe the processes in terms of conformal time $\eta$
\citep{pmf3}. After this procedure the MHD equations include the
effects of the expansion while retaining their conventional flat
spacetime form. To keep the description as simple as possible we
work with dimensionless quantities, such as the normalized
wavenumber\footnote{Here the subscript $\star$ indicates again the
moment of the PMF generation. $\gamma$ can be associated with the
number of PMF bubbles within the Hubble radius, $N \propto
\gamma^3$. This value depends on the PT model: for the QCDPT we
assume $\gamma \simeq 0.15$.} $\gamma =\lambda_0/H^{-1}_\star$ and
normalized energy density defined below.

The coupling between the PMF and the plasma leads to a spreading of
the fixed scale PMF over a wide range of length scales, thus forming
the PMF spectrum. After a few turnover times the modified PMF
spectrum is established (see Sec.~4 for details of the simulations).

To show the coupling between the initial PT-generated
MF and the plasma we give here the basic MHD equations for
an incompressible conducting fluid \citep{B03}
\begin{eqnarray}
\left[\frac{\partial}{\partial\eta} + ({\bm v} \cdot \bm{\nabla} )- \nu
\nabla^2\right] {\bm v} &=& ({\bm b} \cdot \bm{\nabla} ) {\bm b}-\bm{\nabla}
p + {\bm f}_K, \label{mhd1} \\
\left[\frac{\partial}{\partial \eta}+ ({\bm v} \cdot \bm{\nabla} )-\lambda
\nabla^2\right] {\bm b} &=&
({\bm b}\cdot \bm{\nabla} ) {\bm v}
+\bm{\nabla}\times{\bm f}_M, \label{mhd2}
\end{eqnarray}
with $\bm{\nabla}\cdot{\bm b}=0$, where $\eta$ is the conformal
time, ${\bm v}({\bm x},\eta)$ is the fluid velocity, ${\bm b}({\bm
x},\eta) \equiv {\bm B}({\bm x},\eta)/ \sqrt{4\pi w}$ is the
normalized MF, ${\bm f}_K({\bm x},\eta)$ and ${\bm f}_M({\bm
x},\eta)$ are external forces driving the flow and the magnetic
field (${\bm f}_K={\bm f}_M=0$ for the results presented below, but
${\bm f}_M\neq0$ for producing initial conditions), $\nu$ is the
comoving viscosity of the fluid, $\lambda$ is the comoving
resistivity, $w= \rho+p$ is the enthalpy, $\rho$ is the energy
density, and $p$ is the pressure of the plasma. Here we are
interested in the radiation dominated epoch.

To proceed we derive the Fourier transform of the
PMF two point correlation function as
\begin{equation}
\langle b_i^* ({\bm k}, \eta) b_j({\bm k'}, \eta+\tau) \rangle =
(2\pi)^3 \delta({\bm k} -{\bm k'}) \, F_{ij}^M\!({\bm k},\tau) ~.
f[\kappa(k),\tau], \label{2-point}
\end{equation}
Such a presentation allows a direct analogy with hydrodynamic
turbulence \citep{Landau}. In fact, $b_i$ represents the Alfv\'en
velocity. The normalized energy density of the PMF is then
${\mathcal E}_M = \langle {\bm b}^2\rangle/2$, while the
kinetic energy density is ${\mathcal E}_K = \langle {\bm v}^2 \rangle /2$,
and the spectral correlation tensor is
\begin{equation} \frac{
F_{ij}^M\!({\bm k},\tau) }{(2\pi)^3}= P_{ij}({\bm k})
\frac{E_M(k,\tau)}{4\pi k^2} + i \varepsilon_{ijl} {k_l}
\frac{H_M(k,\tau)}{8\pi k^2} \, . \label{eq:4.1}
\end{equation}
Here, $P_{ij}({\bm k}) = \delta_{ij} - {k_i k_j}/{k^2}$ is the
projection operator, $\delta_{ij}$ is the Kronecker delta, $k =
|{\bm k}|$, $\varepsilon_{ijl}$ is the totally antisymmetric tensor,
and $\kappa(k)$ is an autocorrelation function that determines the
characteristic function $f[\kappa(k),\tau]$ describing the temporal
decorrelation of turbulent fluctuations. The function $H_M(k,\eta)$
is the magnetic helicity spectrum. Note that $E_M(k) = k^2
P_B(k)/\pi^2$, where $P_B(k)$ is the MF power spectrum.

The power spectra of magnetic energy $E_M(k,\eta)$ and magnetic
helicity $H_M(k,\eta)$ are related to magnetic energy density and
helicity density through $ {\mathcal E}_M(\eta) = \int^\infty_0 {\rm
d}k E_M(k,\eta) $ and $ {\mathcal H}_M(\eta) = \int^\infty_0  {\rm
d}k H_M(k,\eta)$, respectively. The magnetic correlation length,
\beq \xi_M(\eta) = {1\over{{\mathcal E}_M(\eta)}} {\int^\infty_0
{\rm d}k \, k^{-1} E_M(k,\eta)}, \eeq corresponds to the largest
eddy length scale. All configurations of the MF must satisfy the
``realizability condition'' \citep{B03} \beq |{\mathcal H}_M(\eta)|
\leq 2 \xi_M(\eta) {\mathcal E}_M(\eta). \label{realizability} \eeq
Also, the velocity energy density spectrum $E_K(k,\eta)$ is related
to the kinetic energy of the turbulent motions through ${\mathcal
E}_K(\eta) = \int^\infty_0 {\rm d}k \, E_K(k,\eta)$.

One of the main characteristics of the PMF is the correlation length
and its growth. The maximal correlation length $\xi_{\rm max}$ for a
causally generated PMF cannot exceed the Hubble radius\footnote{The
inflation generated PMF \citep{turner,ratra} correlation length can
exceed the Hubble horizon today.}  at the time of generation
$H_\star^{-1}$. The comoving length corresponding to the Hubble
radius at generation is inversely proportional to the temperature
$T_\star$,
\begin{equation}
\lambda_{H_\star} = 5.3 \times 10^{-7}~{\rm
Mpc}\left(\frac{0.15\,{\rm GeV}}{T_\star}\right)
\left(\frac{15}{g_\star}\right)^{{1}/{6}}, \label{lambda-max}
\end{equation}
and is equal to 0.5 pc for the QCDPT with $g_\star=15$ and
$T_\star=0.15$ GeV.

The PMF spectrum is characterized not only by its spatial
distribution, but also by its characteristic times: (i) the
largest-size eddy turnover time $\tau_0 \simeq l_0/v_A$ (where
$v_A$ is the r.m.s. Alfv\'en velocity), which can also be used to
determine the minimal duration of the source needed to justify the use
of the stationary turbulence approximation \citep{P52,P52-2}; (ii) the
direct cascade time-scale of the turbulence $\tau_{\rm dc}$; and (iii) the
large-scale turbulence decay time $\tau_{\rm ls}$.

The temporal characteristics of the MHD turbulence is given through
the form of $f(\kappa(k),\tau)$, which is due to the complex process of
MHD turbulence decorrelation \citep{ts07} and is currently not
fully understood. To proceed we employ Kraichnan's approach
\citep{K64} and specify the decorrelation function $f_{\rm
dc}[\kappa(k_{\rm ph}),\tau] = \exp \! \left[-\pi \kappa^2(k_{\rm
ph}) \tau^2/4 \right]$ defined within the inertial range,
$k_0<k<k_d$. Here $\tau$ is the duration of the turbulence process
and $\kappa(k_{\rm ph}) = {{\bar \varepsilon}_M^{1/3}}k_{\rm
ph}^{2/3}/{\sqrt{2\pi}}$, where $k_{\rm ph}$ is the physical
wavenumber related to the comoving $k$ through $k_{\rm ph}(a) =
ka_0/a$ ($a_0$ is the value of the scale factor now), and
${\bar\varepsilon}_M$ is the proper dissipation rate per unit
enthalpy. Hence, we have \citep{ktr2011}
\begin{equation}
f_{\rm dc}[{\bar k},\tau] = \exp \! \left[-\frac{ 2 \pi^2}{9} \left(
\frac{\tau}{\tau_0} \right)^2 {\bar k}^{4/3} \right] ~. \label{dec}
\end{equation}
Here, ${\bar k}= k/k_0$ is the normalized wavenumber and $\tau_0$
corresponds to the largest eddy turnover time. It is clear that
after switching off the forcing, the turbulent motions are decorrelated
within a few turnover times, and are in fact irrelevant to influence
the large-scale MF.

\section{Growth of correlation length in helical turbulence}

To assess the importance of a small initial magnetic helicity,
we perform direct numerical simulations of decaying MHD turbulence
with an initial MF of finite relative magnetic helicity using
different values, and a correlation length $\xi_M$ that is small
compared with the scale of the domain $\lambda_1$.

\subsection{Simulation technique}

We solve the compressible equations with the pressure given by
$p=\rho c_s^2$, where $c_s=1/\sqrt{3}$ is the sound speed for an
ultra-relativistic gas. Following our earlier work \citep{kbtr10},
we solve the equations governing equations for the logarithmic
density $\ln\rho$, the velocity $\bm{v}$, and the magnetic vector
potential $\bm{A}$, in the form
\begin{eqnarray}
\frac{\DD\ln\rho}{\DD\eta}&=&- \bm{\nabla}\cdot\bm{v},\\
\frac{\DD\bm{v}}{\DD\eta}&=&\bm{J}\times\bm{B}-c_s^2\bm{\nabla}\ln\rho
+\bm{f}_{\rm visc},
\\
\frac{\partial\bm{A}}{\partial \eta}&=&
\bm{v}\times\bm{B}
+{\bm f}_M+\lambda \nabla^2{\bm A},
\label{mhd3}
\end{eqnarray}
where $\DD/\DD\eta=\partial/\partial\eta+\bm{v}\cdot\bm{\nabla}$
is the advective derivative, $\bm{f}_{\rm visc}=\nu\left(\nabla^2{\bm v}
+{\textstyle\frac{1}{3}}\bm{\nabla}\bm{\nabla}\cdot\bm{v}+\bm{G}\right)$
is the viscous force in the compressible case with constant $\nu$
and $G_i={\sf S}_{ij}\nabla_j\ln\rho$ as well as
${\sf S}_{ij}=\frac{1}{2}(v_{i,j}+v_{j,i})-\frac{1}{3}\delta_{ij}v_{k,k}$
being the trace-free rate of strain tensor.
Furthermore, $\bm{J}=\bm{\nabla}\times\bm{B}/4\pi$ is the normalized
current density.
We emphasize that ${\bm f}_M=\bm{0}$, except for producing initial
conditions, as explained below.

The bulk motions are always slow enough, so compressibility effects
are not important. Similar to before, we express the magnetic field
in Alfv\'en units, but now based on the volume average enthalpy,
i.e., ${\bm b}\equiv {\bm B}/\sqrt{4\pi\langle w\rangle}$, where
$w={4\over3}\rho$ for an ultrarelativistic gas. We use $512^3$
meshpoints in a domain of size $(2\pi)^3$, so the lowest wavenumber
in the domain is $k_1=1$. We choose $\nu=\eta=10^{-5}$ in units of
$c_s/k_1$.

\subsection{Initial conditions}

A suitable initial condition is produced by simulating
for a short time interval ($\Delta t\approx0.5 \lambda_1/c_s$)
with a random $\delta$-correlated
magnetic forcing term $\ff_M$ in the evolution equation for the
magnetic vector potential. This forcing term consists of plane
monochromatic waves with wavenumber $k_0$ and fractional helicity
$\bra{\ff_M\cdot\nab\times\ff_M}/\bra{k_0\ff_M^2}=2\sigma/(1+\sigma^2)$;
in the following we quote the value of $\sigma$. This
procedure has the advantage that the magnetic and velocity fields used
then for the subsequent decay calculations are obtained from a
self-consistent solution to the MHD equations.

\subsection{Growth of helical structures}

\begin{figure}[t]
\begin{center}
\includegraphics[width=0.92 \columnwidth]{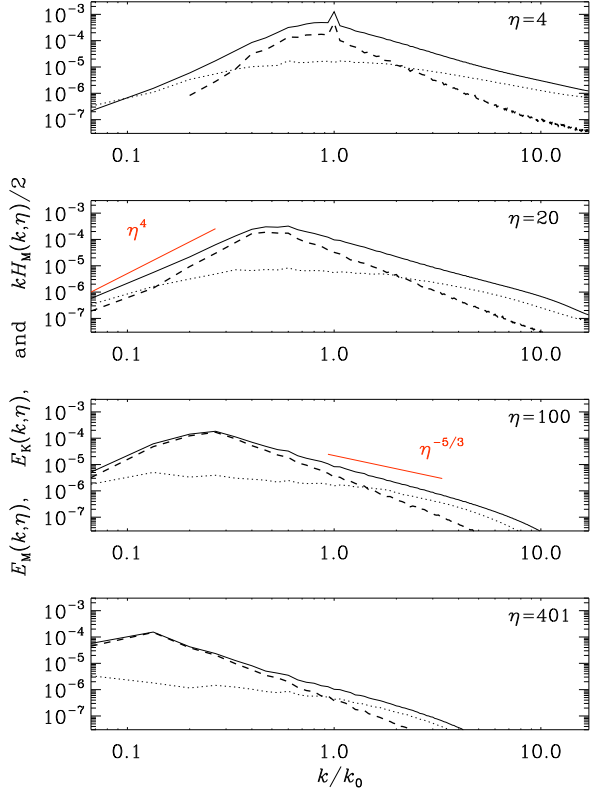}
\end{center}\caption[]{
Spectra of magnetic energy (solid lines), kinetic energy (dotted),
and magnetic helicity scaled by $k/2$ (dashed) for a run with $\sigma=0.03$
at three different times.
At early times, $H_M(k,\eta)$ can be negative at small values
of $k$, which explains why the dashed line terminates in those cases.
}\label{short512pm1a3_sig003_noforce2}
\end{figure}

In \Fig{short512pm1a3_sig003_noforce2} we show spectra of magnetic
and kinetic energy, as well as the magnetic helicity scaled by
$k/2$, for a run with $\sigma=0.03$. Initially, $kH_{\rm
M}(k,\eta)/2$ is well below the value of $E_M(k,\eta)$.
However, at later times the two approach each other at large scales. This shows
that the relative magnetic helicity increases during the decay.
For the four times shown in \Fig{short512pm1a3_sig003_noforce2},
the rms Mach number, $v_{\rm rms}/c_s$, is 0.05, 0.025, 0.012, and 0.007;
$b_{\rm rms}/v_{\rm rms}$ is around 3.4, and the Reynolds numbers
$v_{\rm rms}\xi_M/\nu$ are roughly 270 all cases.

\begin{figure*}[t]
\begin{center}
\includegraphics[width=\textwidth]{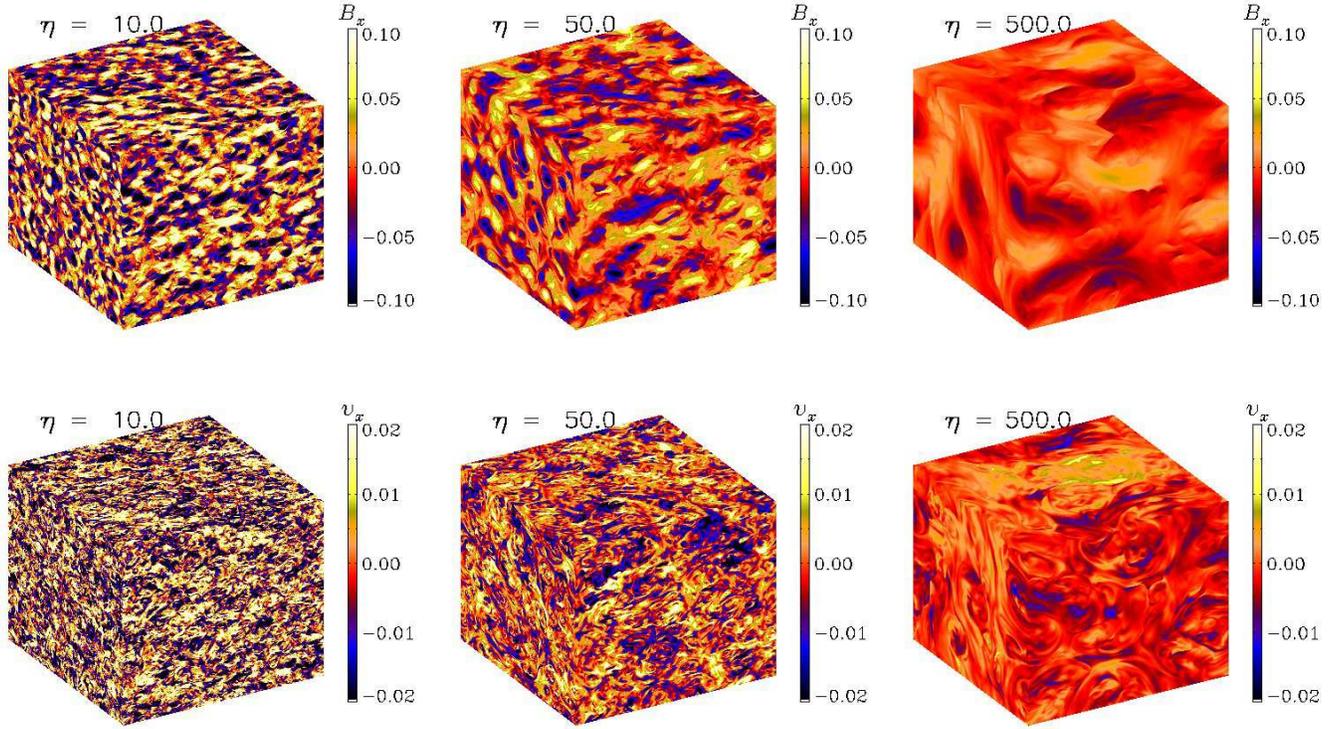}
\end{center}\caption[]{
Visualizations of $B_x$ (upper row) and $v_x$ (lower row) at three
times during the magnetic decay of a weakly helical field with
$\sigma=0.03$.}\label{BU}
\end{figure*}

The growth of turbulent structures is particularly clear in the
magnetic field (\Fig{BU}).
The magnetic field drives correspondingly larger scale structures also in the
velocity field.
However, there are also strong small-scale fluctuations
in the velocity field that are not visible in the
magnetic field; see the second row of \Fig{BU}.

In agreement with earlier simulations, we find that at small scales
the magnetic energy is re-distributed by a direct cascade with a
Kolmogorov-type spectrum, $E_M(k) \propto k^{-5/3}$. At large scales
a Batchelor spectrum\footnote{Sometimes this spectral distribution,
$E_M(k) \propto k^4$, is called a von K\'arm\'an spectrum
\citep{pope}.}, $E_M(k) \propto k^4$, is established, which was
used as initial condition already in \cite{pmf3}. This spectrum is in
agreement with the analytical description of
\cite{caprini-1} who derived this result from the requirement of
causality and the divergence free condition. The earlier study of
\cite{hogan}, which thus violates causality for magnetic energy,
yielded a white noise
spectrum $E(k) \propto k^2$ (Saffman spectrum) which we do
observe for the spectral distribution of the kinetic energy $E_K(k)
\propto k^2$.

\begin{figure}[t]
\begin{center}
\includegraphics[width=\columnwidth]{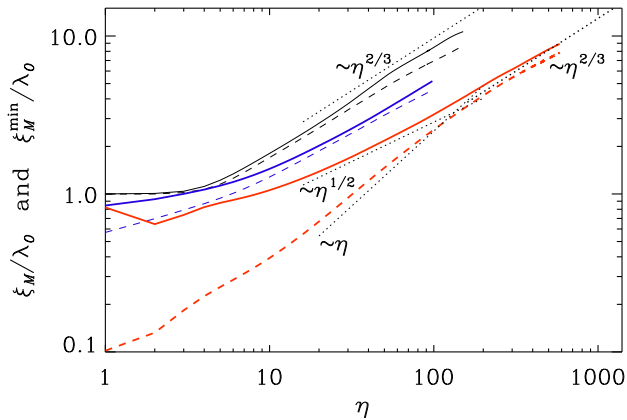}
\end{center}\caption[]{
Evolution of $\xi_M(\eta)$ (solid) and $\xi_M^{\min}(\eta)$ (dashed)
for $\sigma=1$ (black), 0.1 (blue) and 0.03 (red).
}\label{pcomp_kft_QCD}
\end{figure}

\subsection{Growth of turbulent length scales}

The evolution of magnetic correlation length and magnetic energy
during the MHD turbulence decay can be described using two indices
$n_{\xi}$ and $n_E$:
\begin{equation}
\xi_M(\eta) = \xi_M(\eta_0) \left({\eta \over \eta_0}\right)^{n_\xi} ~,
\label{xi(eta)}
\end{equation}
\begin{equation}
{\mathcal E}_M(\eta) = {\mathcal E}_M(\eta_0) \left({\eta \over \eta_0}\right)^{n_E} ~.
\label{E(eta)}
\end{equation}
In this case, we can model the spectral energy density of the PMF
using time-dependent large- and small-scale ranges:
\begin{equation}
E_M(k,\eta) = E_0(\eta)
\left\{ \begin{array}{l}
\bar k^4 ~~~~~~{\rm when}~~ k<k_I(\eta)  \\ \bar k^{-5/3} ~~{\rm when}~~ k>k_I(\eta)
\end{array} \right. ~,
\label{spectrum}
\end{equation}
where $\bar k = k / k_I$ and $k_I(\eta) = 2 \pi / \xi_M(\eta)$.
Hence, the evolution of the spectral amplitude $E_0$ for a given
magnetic field spectrum will be [see Equations~(\ref{xi(eta)})
and (\ref{E(eta)})]:
\begin{equation}
E_0(\eta) = {5 \over 17 \pi} \xi_M(\eta_0) {\mathcal E}_M(\eta_0)
\left({\eta \over \eta_0}\right)^{n_\xi + n_E} ~.
\label{E0(eta)}
\end{equation}

Magnetic helicity crucially affects the evolution of the PMF
\citep{BN99,BN991,BN992,Axel-1,axel,jedamzik1,jedamzik2,campanelli}.
If the PMF has been generated with small magnetic helicity, there
are two main stages during the development of the MF spectrum: during
the first stage (sometimes called direct cascade) the PMF dynamics
is similar to that of the non-helical MF. The energy cascades from
large to small scales where it decorrelates and dissipates: this
is a standard forward cascade development. Since magnetic helicity
is conserved, its fractional value increases and thus the end of
this first stage is characterized by releasing turbulence to a
maximally helical state \citep{jedamzik1,axel} when the realizability
condition (\ref{realizability}) is reached, the inverse-cascade stage starts.
The conservation of magnetic helicity implies that the magnetic energy
density decays in inverse proportion to the correlation length
growth during the inverse cascade. The realizability condition implies that
\begin{equation}
\xi_M(\eta)\ge\xi_M^{\min}(\eta)\equiv|{\mathcal H}_M(\eta)|/2{\mathcal E}_M(\eta),
\end{equation}
so there is a minimum value for the correlation length. In
\Fig{pcomp_kft_QCD} we plot $\xi_M(\eta)$ and $\xi_M^{\min}(\eta)$
for $\sigma=1$, 0.1, and 0.03. It turns out that, especially in the
latter case with $\sigma=0.03$, the increase of $\xi_M$ remains slow
($\sim\eta^{1/2}$) as long as $\xi_M(\eta)\gg\xi_M^{\min}(\eta)$.
However, since ${\mathcal H}_M$ is essentially constant and
${\mathcal E}_M$ decreases approximately like $\eta^{-1}$, the value
of $\xi_M^{\min}(\eta)$ soon reaches $\xi_M(\eta)$. When that
happens, the field is essentially fully helical and the correlation
length and the magnetic energy density evolve according to
$\xi_M\sim\eta^{2/3}$ and ${\mathcal E}_M\sim\eta^{-2/3}$,
respectively. Hence, we recover two distinctive phases in the MHD
turbulence decay process: evolution of a \emph{weakly helical}
turbulence with $n_{\xi} = 1/2$ and $n_E = -1$, and \emph{fully
helical} turbulence with $n_{\xi} = 2/3$ and $n_E = -2/3$. Note
that in the latter case $E_0(\eta) \propto \xi_M(\eta) {\mathcal
E}_M(\eta) = {\rm const}$ [see Eq.~(\ref{E0(eta)})] and the inverse
cascade develops. Our results are in excellent agreement with
\cite{BN99}, \cite{BN991}, \cite{jedamzik1}, and
\cite{campanelli}. The dynamical process of
PMF coupling with the cosmic plasma stops at the moment of
recombination after which the PMF develops more slowly \citep{BEO97}.

To calculate the time $\eta_{\rm fully}$ when a fully helical state
is reached, we only need to know the initial values $\xi_M(\eta_0)$
and $\xi_M^{\min}(\eta_0)$.
Since the latter approaches the former like $\eta^{1/2}$, the result
is $\eta_{\rm fully}=\eta_0[\xi_M(\eta_0)/\xi_M^{\min}(\eta_0)]^2$.
Thus, in terms of the initial values of ${\mathcal E}_M$ and
${\mathcal H}_M$, a fully helical state is reached at the time
\begin{equation}
\eta_{\rm fully}=4\eta_0 \xi_M^2 {\mathcal E}_M^2/{\mathcal H}_M^2.
\end{equation}
Note that this time increases quadratically with decreasing initial
value of ${\mathcal H}_M$. In case of the strong CP violation during
the QCDPT, when the initial magnetic helicity can reach values that
are only $\xi_M / \lambda_{H_\star}$ times less than the maximal one
(see Sec.~2), we get $\eta_{\rm fully} = \eta_0 / \gamma^2 $.

\begin{figure}[t]
\begin{center}
\includegraphics[width=0.99\columnwidth]{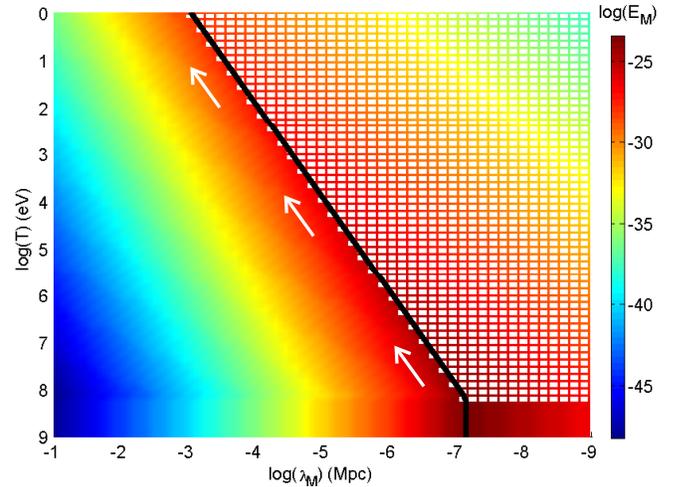}
\end{center}\caption[]{
Spectral energy density of the turbulent magnetic field
log$E_M(k)$ (color coded) in a representation of magnetic
correlation length versus temperature.
The thick solid line shows the evolution of the magnetic
correlation length $\xi_M(T)$.
The magnetic correlation length starts to grow after the QCD
phase transition at $T_\star = 0.15$ GeV, when $\xi_M = 0.075$ pc.
The transparent dashed area corresponds to decorrelated magnetic
field. White arrows show the direction of the evolution during
the expansion of the Universe. Here $n_{\xi}=-1/2$ and $n_{E}=1$.
}\label{Emk_weakly}
\end{figure}

\begin{figure}[t]
\begin{center}
\includegraphics[width=0.99\columnwidth]{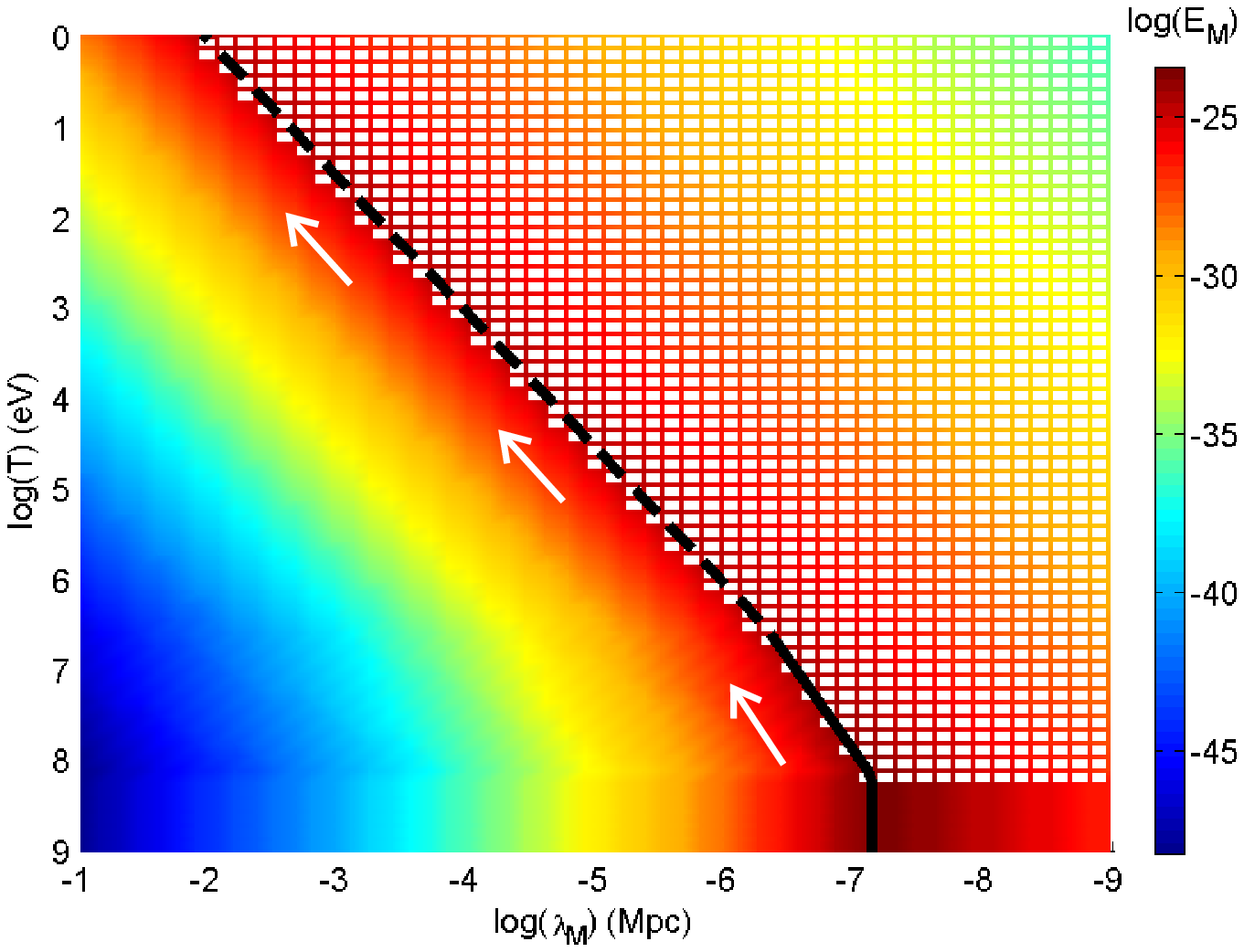}
\end{center}\caption[]{
Similar to Fig.~\ref{Emk_weakly}, but for the case in which the magnetic
field reaches a fully helical state within the considered expansion
time of the Universe. Initially, during the
growth of magnetic helicity correlation length (solid line) and
energy evolve according to $n_{\xi}=-1/2$ and $n_{E}=1$.
After reaching a fully helical state, correlation length (dashed line)
and energy evolve according to $n_{\xi}=-2/3$ and
$n_{E}=2/3$.}\label{Emk_fully}
\end{figure}

\subsection{Observed Magnetic Fields}

Galactic and cluster MFs are usually measured through
Faraday rotation \citep[see][]{Vallee} and, as mentioned above,
the value of the coherent magnetic field is of the order of
a few $\mu$G with a typical
coherence scale of 10\,kpc,\footnote{Strong MFs have been
detected through Faraday rotation of distant quasars
proving that the MFs comparable to those observed today are seen at
high redshift $z \sim 3$ \citep{nature}.} and cluster MFs have
lower limits of the order of $10^{-6}$\,G, and at least a few nG, with
similar coherence scales \citep{Clarke} and
additional lower limits on the steepness of the magnetic power spectrum
in clusters. Furthermore, simulations starting from a
constant comoving magnetic fields of $10^{-11}$\,G suggest that magnetic
field generation in clusters can be sufficiently strong to explain
Faraday rotation measurements \citep{Dolag,Jedamzik:2010cy}.

Figure~\ref{Emk_weakly} shows the spectral energy density of the
QCDPT-generated MF \citep[see][]{kisslinger} with respect to temperature
and correlation length in weakly helical turbulence. Initially
the integral scale of the MHD turbulence is set by the QCDPT bubble
scale (lower right corner of the diagram). The thick solid line
marks the division between the evolution of large-scale (plain
colored region) and small-scale (hashed colored region) magnetic
fields. White arrows indicate the direction of the evolution during
the Universe expansion. At scales below the integral scale of the
turbulence, the magnetic field undergoes exponential decorrelation;
see Eq.~(\ref{dec}). The integral scale of the MHD turbulence
increases, reaching $\xi_M = 1$\,kpc at $T = 1$\,eV. Here we
have used $n_\xi = 1/2$, $n_E = -1$, with initial magnetic helicity
corresponding to that given by Eq. (\ref{weakhel}).

Figure~\ref{Emk_fully} shows the spectral energy density of a
QCDPT-generated MF in the case when the initial helical turbulence reaches
the fully helical case during the expansion of the Universe with $\eta_{\rm
fully}/\eta_0 = 1/(0.15)^2$. The thick solid line marks the
evolution of the magnetic correlation length until the magnetic helicity reaches
its maximally allowed value. In this time interval the decay law for weakly
helical turbulence with $n_\xi = 1/2$ and $n_E = -1$ is applied. After
the time $\eta_{\rm fully}$ when maximal magnetic helicity is
reached, the correlation length follows the black dashed line and the MF
evolution follows that of the fully helical case with $n_\xi = 2/3$ and $n_E =
-2/3$. The integral scale of the MHD turbulence reaches $\xi_M = 10$\,kpc
at $T = 1$\,eV.

The presented model is somewhat idealized since it ignores the time of Silk
damping due to large correlation lengths for photon and neutrino
viscosity \citep[see][]{Jedamzik:2010cy}. This is justified since it
only delays the evolution but does not destroy the field \citep{BEO97}.
Therefore, we can present here only upper values for QCDPT MFs
within the model by \cite{kisslinger}.

The final amplitude of the MF can be estimated through two different
approaches. (i) We compute the total magnetic energy density, i.e.\
${\mathcal E}_M = \int_0^\infty {\rm d}k\, E_M(k)$ and make the assumption
that all energy is again given only at {\it one} scale that
corresponds to the integral scale at this moment, i.e.\ $B_{\rm
eff}=\sqrt{8\pi{\mathcal E}_M}$. (ii) Another approach is to compute
the strength of the MF, $B(\lambda)$, at a given scale $\lambda$.
Since observations \citep{Vallee} do not allow us to properly
reconstruct the configuration of the MF we adopt first an
``effective'' MF approach \citep[see][]{ktr2011}. The resulting
value of the effective MF in our model of weakly helical
turbulence with $\xi_M = 1$\,kpc reaches $5 \times 10^{-4}\,{\rm nG}$, while in the case of
a fully helical configuration with $\xi_M = 10$\,kpc we find $7 \times 10^{-3}\,{\rm nG}$.

\section{Conclusions}

In this paper we have considered QCDPT-generated PMFs and their
evolution in an expanding Universe accounting for the effects of
MHD turbulence to explain the seed MFs of clusters and
galaxies. We consider the MF generation model proposed by
\cite{kisslinger}, which yields an initial state of weakly helical
MHD turbulence. We also study the possibility of strong CP
violation according to \cite{fz00}, which yields an initial state
with much higher magnetic helicity at a time when maximal helicity of the MHD
turbulence is reached during the expansion of the Universe. The initial
seed MF is generated via QCDPT bubble collisions with a comoving
correlation length of the order of 0.1 pc and with a comoving
amplitude of the order of 20\,nG. The initial magnetic helicity is
determined by the thickness of the surface between two colliding bubbles and
is extremely small if no strong CP violation is assumed
\citep{cp2,cp3,cp4}. During the expansion of the Universe there are
different processes that affect  the correlation length and the
strength of the MF: first of all, during the PT the field is
initially peaked at a given scale and then spreads out within a wide
range of wavelengths, establishing a Kolmogorov-like spectrum,
$E_M(k) \propto k^{-5/3}$, at small scales and a Batchelor spectrum,
$E_M(k) \propto k^4$, at large scales. If the PMF was generated
without being maximally helical, the magnetic helicity experiences a
steady growth. One of the results obtained in this paper is an
estimate of the timescale within which the field starts to be fully
helical. In the case of an extremely weakly helical field
\citep{kisslinger}, the available time to produce a fully helical
PMF may be too long. The growth of the correlation length follows
then a $\xi_M \propto T^{-1/2}$ law. For moderate or reasonably
small initial magnetic helicity (even for $\sigma \geq
10^{-6}-10^{-5}$), the evolution timescale is long enough so that
during the first stage of evolution, magnetic helicity grows to its
maximal value. During the next stages (after magnetic helicity has
reached its maximal value) the correlation length experiences a
steady growth with the scaling law $\xi_M \propto T^{-2/3}$  while
the energy density is decreasing in the opposite way keeping
magnetic helicity almost constant. Finally, at recombination the
growth of the correlation length slows down.
The resulting correlation length in the
most optimistic scenarios is around 10\,kpc and the amplitude of the
MF is around 0.007\,nG. Assuming that the MF is amplified during the
growth of structures \citep{Dolag}, such a field might well be
strong enough to explain the observed MF in galaxies and clusters.
On the other hand, observations of the CMB fluctuations are
sensitive to PMFs of the order of a few nG \citep[see][and
references therein]{shaw,yamazaki}.

Another possible signature of QCDPT-generated magnetic fields is a
gravitational wave signal \citep{kks10} that might be indirectly
detected through pulsar timing \citep{Durrer3}. The gravitational
waves signal from PTs is usually computed assuming short duration of
the source (either turbulence or PMF anisotropic stress). On the
other hand, due to the free decay of MHD turbulence, the source of
gravitational waves acts also after the end of PTs. For short
duration sources, the peak frequency of the gravitational waves is
fully determined by the source characteristics. In particular, for
QCDPT-generated gravitational waves it is far too weak to be
detected though gravitational waves via ground or space based
missions. Long duration sources might in principle substantially
change the peak frequency as well as the amplitude of the signal. We
plan to address this issue in future work.

\acknowledgments  We appreciate helpful comments and discussions
with L. Campanelli, C.\ Caprini, R.\ Durrer, A.\ Kosowsky, K, Kunze,
A. Neronov, B. Ratra, and T.\ Vachaspati. We acknowledge partial
support from Computing resources have been provided by the Swedish
National Allocations Committee at the Center for Parallel Computers
at the Royal Institute of Technology in Stockholm and by Carnegie
Mellon University supercomputer center. We acknowledge partial
support from Swiss National Science Foundation SCOPES grant no.
128040, NSF grant AST1109180 and NASA Astrophysics Theory Program
grant NNXlOAC85G. This work was supported in part by the European
Research Council under the AstroDyn Research Project 227952 and the
Swedish Research Council grant 621-2007-4064. T.K.\ acknowledges the
ICTP associate membership program. A.B.\ and A.T.\ acknowledge the
McWilliams Center for Cosmology for hospitality.

\newcommand{\yaraa}[3]{ #1, {ARA\&A,} {#2}, #3}
\newcommand{\yjour}[4]{ #1, {#2}, {#3}, #4}

\end{document}